\documentclass[aps,prl,twocolumn]{revtex4}
\usepackage{graphicx}

\begin{document}

\title{A Novel Dielectric Anomaly in Cuprates and Nickelates: Signature of an Electronic Glassy State}
\author{Tuson Park$^{1}$, Z. Nussinov$^{1}$, K. R. A. Hazzard$^{2}$, V. A. Sidorov$^{1*}$, A. V. Balatsky$^{1}$, J. L. Sarrao$^{1}$, S.-W. Cheong$^{3}$, M. F. Hundley$^{1}$, Jang-Sik Lee$^{1}$, Q. X. Jia$^{1}$, J. D. Thompson$^{1}$}
\affiliation{$^{1}$Los Alamos National Laboratory, Los Alamos, New Mexico 87545 \\$^{2}$ Department of Physics, Ohio State University, 174 West 18th Avenue, Columbus, Ohio 43210 \\$^{3}$ Department of Physics and Astronomy, Rutgers University and Bell Laboratories, Lucent Technologies, Murray Hill, New Jersey 07974}
\date{April 19th, 2004}

\begin{abstract}
The low-frequency dielectric response of hole-doped insulators La$_{2}$Cu$_{1-x}$Li$_{x}$O$_{4}$ and La$_{2-x}$Sr$_{x}$NiO$_{4}$ shows a large dielectric constant $\varepsilon ^{\prime}$ at high temperature and a step-like drop by a factor of 100 at a material-dependent low temperature $T_{f}$. $T_{f}$ increases with frequency and the dielectric response shows universal scaling in a Cole-Cole plot, suggesting that a charge glass state is realized both in the cuprates and in the nickelates.
\end{abstract}

\maketitle
Charge inhomogeneities in hole-doped oxides attract great interest, in part due to their possible relation to high temperature superconductivity. Perhaps the best known examples are stripes, wherein holes congregate along lines which serve as domain boundaries in a surrounding antiferromagnetic environment. These were predicted \cite{zaanen89,emery93} and observed in Nd-doped La$_{2-x}$Sr$_{x}$CuO$_{4}$ \cite{tranquada95} and in hole-doped, but insulating nickelates \cite{chen93,tranquada94,sachan95}. Inhomogeneities in these complex oxides are naturally linked to the competition between myriad interactions (e.g. Coulomb, kinetic, magnetic, strain,...). The resulting compromise amongst these is non-uniform structures \cite{zaanen89,emery93} and experimental evidence for glassiness \cite{klauss00,julien99}. Theoretically, the non-uniform nature of these states easily allows for the proliferation of many metastable and low lying states. In accord with this intuition, replica calculations  \cite{schmalian00,wu03} indeed suggest an exponentially large number of metastable states built from these rich non-uniform domains. This proliferation of metastable states allows, in principle, for a glass to naturally emerge. As spin and charge degrees of freedom are intertwined, glassy signatures and inhomogeneities in the spin sector naturally suggest analogous effects in the charge channel- nonuniform charge distributions should appear.

In this Letter, we report the first observation of sharp dielectric anomalies  both in the cuprates and nickelates. The low-frequency dielectric constants of La$_{2}$Cu$_{1-x}$Li$_{x}$O$_{4}$ ($x\approx 0.023$) and La$_{2-x}$Sr$_{x}$NiO$_{4}$ ($x=1/3$) show a frequency-dependent drop from a large dielectric constant $\varepsilon ^{\prime}$, of order $10^{4}$, at high temperatures to a modest value ($10^{2}$) below a temperature $T_{f}$. A phenomenological Davidson-Cole plot \cite{davidson51}, commonly used to analyze classical structural glasses \cite{davidson51,menon95},  successfully captures the dielectric behavior which shows universal scaling. The characteristic relaxation time obtained from the analysis shows a polynomial temperature dependence with a finite glass transition temperature $T_{0}$, which is consistent with glassy charge dynamics arising from a collective behavior of doped holes both in the cuprates and in the nickelates.

The Li-doped lanthanum cuprate La$_{2}$Cu$_{1-x}$Li$_{x}$O$_{4}$ crystals were prepared by flux growth and the Sr-doped lanthanum nickelate La$_{2-x}$Sr$_{x}$NiO$_{4}$ was grown by the floating zone method \cite{lee97}. For dielectric measurements, an Al$_{2}$O$_{3}$ layer of 1000~$\AA$ was deposited between the sample and metal contacts to prevent a polarization of charge at an electrode/sample interface that can produce a spurious dielectric response. Extensive tests were made to ensure that the intrinsic reproducible response of these crystals was determined. The complex dielectric constant was measured with a precision LCR meter (Agilent 4284~A) from 100 Hz to 1 MHz and the contribution from Al$_{2}$O$_{3}$ subtracted. 

\begin{figure}[tbp]
\centering  \includegraphics[width=8.6cm,clip]{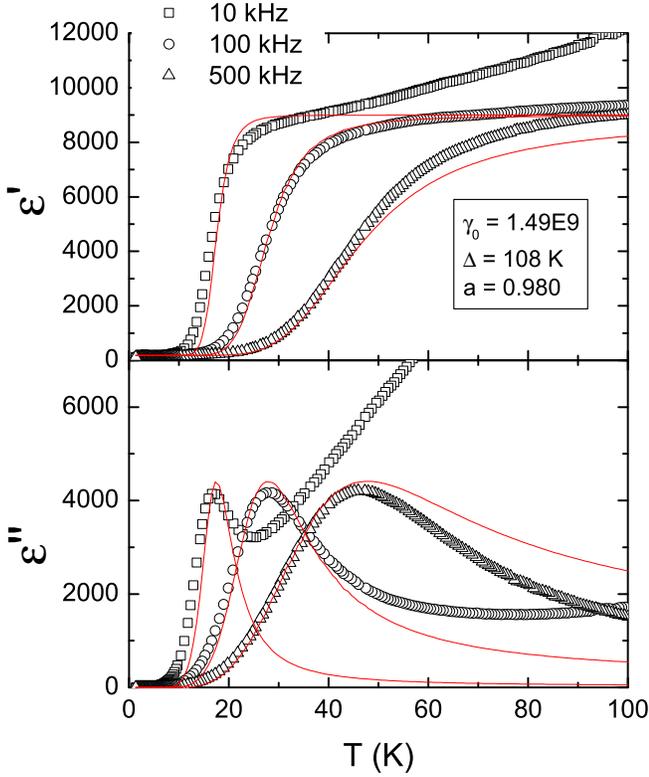}
\caption{The complex dielectric constant of La$_{2}$Cu$_{1-x}$Li$_{x}$O$_{4}$ ($x=0.023$) is shown against temperature, i.e., the real part $\varepsilon ^{\prime}$ in the top panel and the imaginary part $\varepsilon ^{\prime \prime}$ in the bottom panel. Solid lines are results from defect model (see text).}
\label{figure1}
\end{figure} 
The complex dielectric constant of a representative La$_{2}$Cu$_{1-x}$Li$_{x}$O$_{4}$ crystal with ($x\approx 0.023$) is shown as a function of temperature at several frequencies in Fig.~1. The real part $\varepsilon ^{\prime}$ reveals a pronounced step-like increase from of order $10^{2}$ to 10$^{4}$ at $T_{f}$ for a measuring frequency of 10 kHz. $T_{f}$ increases with frequency, which is reminiscent of relaxor-like behavior. The imaginary part $\varepsilon ^{\prime \prime}$ also shows a corresponding peak at $T_{f}(\omega)$. Very similar responses in $\varepsilon ^{\prime}$ and $\varepsilon ^{\prime \prime}$ also were found for crystals with $0.01<x<0.04$. A comparable frequency-dependent drop in dielectric constant has been reported in CaCu$_{3}$Ti$_{4}$O$_{12}$ \cite{subramanian00} and a defect model proposed as a possible explanation \cite{ramirez02}. In the model,  isolated defects such as Cu vacancies produce a local disruption of the ideal cubic structure. The defective regions, then, undergo distortions characteristic of the perovskite structure and relax between alternate equivalent configurations preserving, on average, its global crystal structure. The relaxation rate $\gamma$ is determined by an energy barrier $\Delta$ between alternate configurations, \textit{i.e.}, $\gamma = \gamma _{0} \exp{(-\Delta /T)}$ where $\gamma _{0}$ depends on the effective mass of the defect and is order of $10^{10} - 10^{12}$~Hz. In a mean-field approximation, the dielectric function of a lattice with isolated defects is given by:
\begin{equation}
\varepsilon(\omega,T) = \frac{\varepsilon _{0}}{1-\frac{a \gamma}{-i\omega +\gamma}}.
\end{equation}
Here, $\varepsilon_{0}$ is the dielectric constant at zero temperature and $a$ is a dimensionless parameter which depends on the defect concentration $c$ and the polarizability of the defect $p$, \textit{i.e.}, $a=(4\pi /3)cp\varepsilon _{0}$. We used one representative data set of the real part $\varepsilon ^{\prime}$ of 100~kHz data for the least-square fit of Eq.~(1). The best result was obtained with $a=0.98$, $\gamma _{0}=1.49\times 10^{9}$, and $\Delta = 108$~K. It is interesting to note that the obtained barrier energy $\Delta$ is of the same order of the characteristic energy for the freezing of spin degrees of freedom in the same material \cite{suh98}. The same fitting parameters were used to calculate the temperature dependence of the real part $\varepsilon ^{\prime}$ and the imaginary part $\varepsilon ^{\prime \prime}$ at different frequencies. The solid lines in Fig.~1 are the model calculations, which reproduce the experimental data well. 

\begin{figure}[tbp]
\centering  \includegraphics[width=8.6cm,clip]{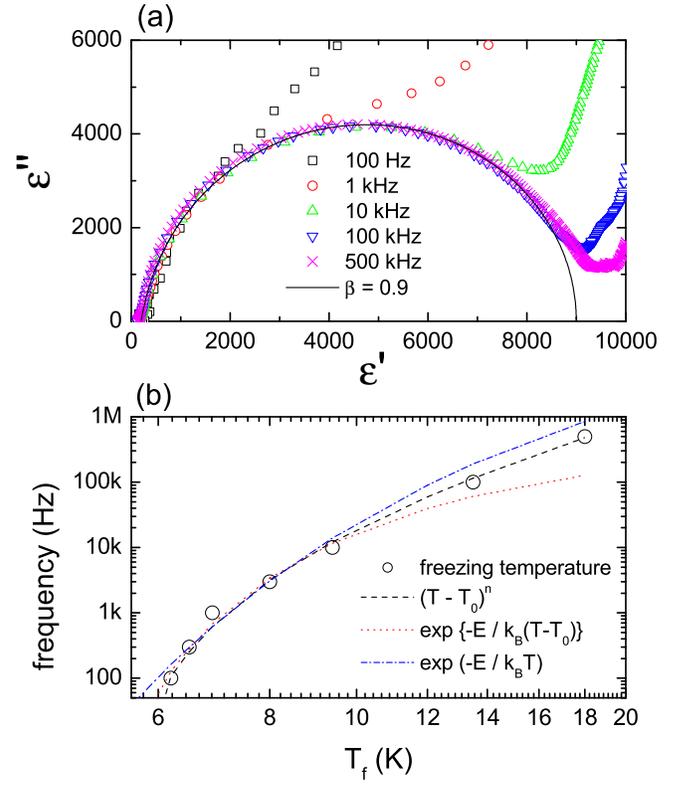}
\caption{(a) The imaginary part $\varepsilon ^{\prime \prime}$ of La$_{2}$Cu$_{1-x}$Li$_{x}$O$_{4}$ ($x=0.023$) is plotted against the real part $\varepsilon ^{\prime}$ for $E \parallel c-$axis. The solid line is the best fit of the Davidson-Cole analysis with $\beta = 0.9$. (b) The measured frequency is plotted against freezing temperature $T_{f}$ obtained from the Cole-Cole arc.}
\label{figure2}
\end{figure}
The sharp drop at low temperature in $\varepsilon ^{\prime}$ and frequency dependence of the transition temperature can be also considered more generally as characteristics of glassy behavior. In Fig.~2a, the imaginary part of the dielectric constant $\varepsilon ^{\prime \prime}$ of La$_{2}$Cu$_{1-x}$Li$_{x}$O$_{4}$ is shown as a function of the real part $\varepsilon ^{\prime}$, namely a Cole-Cole \cite{cole41} or Davidson-Cole plot \cite{davidson51}. The Davidson-Cole arc is given by
\begin{equation}
\frac{\varepsilon - \varepsilon ^{\prime}_{\infty}}{\varepsilon ^{\prime}_{0} - \varepsilon ^{\prime}_{\infty}} = \frac{1}{(1+i\omega \tau _{0})^{\beta}},
\end{equation}
where the complex dielectric function is $\varepsilon = \varepsilon ^{\prime} + i\varepsilon ^{\prime \prime}$ and $\tau _{0}$ is a relaxation time. From fitting to Eq. (2), we determine the zero-frequency dielectric constant $\varepsilon ^{\prime}_{0}$, the limiting high-frequency value $\varepsilon ^{\prime}_{\infty}$, and the skewness $\beta$. When $\beta =1$, the expression reduces to the Debye single relaxation model \cite{debye29}. Deviations from the Debye form point to the multitude of relaxation times present. All of the data at several frequencies collapse on the calculated arc with $\beta =0.9$ (solid line in Fig.~2a), showing universal scaling behavior in the dielectric response. We note that data at low frequency deviate from the calculated arc at $\varepsilon ^{\prime} \approx 2000$, while the data at higher frequency deviate at larger $\varepsilon ^{\prime}$. These deviations are connected with the divergence of $\varepsilon ^{\prime \prime}$ at high temperatures (see Fig.~1). As the contribution from mobile charge carriers depends on the \textit{dc} electrical conductivity $\sigma _{dc}$ and the measuring frequency, $\varepsilon ^{\prime \prime}_{dc} = \sigma _{dc}/ \omega \varepsilon _{0}$, the divergence is more prominent at low frequency and high temperature. The resistivity of the cuprate at room temperature is $\approx 400$~m$\Omega$ cm and exceed 4~$\Omega$ cm below 100~K.

The glass-like charge behavior in the cuprate may be considered as an electrical analog of a glass transition for spins. The randomly oriented dipoles will be frozen at a freezing temperature $T_{f}$, resulting in divergence of the relaxation time ($\omega \tau _{0} \rightarrow \infty$). In polar coordinates $(r,\theta)$ with $r = \sqrt{(\varepsilon^{\prime} - \varepsilon^{\prime}_{\infty})^{2} +
(\varepsilon^{\prime \prime})^{2}}$ and $\theta= \tan^{-1} [\varepsilon^{\prime 
\prime}/(\varepsilon^{\prime} - \varepsilon^{\prime}_{\infty})]$, the Davidson-Cole expression can be written as\\
\begin{equation}
\begin{array}{ll}
r = (\varepsilon ^{\prime}_{0}-\varepsilon ^{\prime}_{\infty})[\cos{(\theta /\beta)}]^{\beta},\\
\omega \tau _{0} = \tan{(\theta / \beta)}.
\end{array}
\end{equation}
As $\omega \tau _{0} \rightarrow \infty$, the data should fall on a straight line intercepting the horizontal $\varepsilon ^{\prime}$ axis at $\varepsilon ^{\prime}_{\infty}$ and at angle of $\theta = \beta \pi /2$. The freezing temperature $T_{f}$ can be identified as the highest temperature at which the points near the $\varepsilon ^{\prime} =\varepsilon ^{\prime}_{\infty}$ intercept (on the left of the arc in Fig.~2) are on a straight line. Fig.~2b shows the frequency change as a function of freezing temperature of La$_{2}$Cu$_{1-x}$Li$_{x}$O$_{4}$ on a log-log scale. There is a clear downward curvature, indicating the zero-frequency freezing temperature $(T_{0})$ is not zero. The inadequacy of an activated or Arrhenius behavior of the relaxation rates, $\tau(T) = \exp{(E/k_{B}T)}$, can be also seen in Fig.~2b (dash-dotted line), suggesting that a freezing of hopping electrons \cite{ikeda94} may not be the origin of the relaxational behavior observed in this material. In spin-glass transitions and within the dynamical mode coupling schemes for supercooled liquids, a critical slowing down of the relaxation rates is generic, $\tau (T) = a(T-T_{0})^{-n}$, where $n$ depends on the particular system \cite{edwards75}. In contrast, the relaxation times may, a priori, follow a more dramatic Vogel-Fulcher (VF) increase commonly observed in glassy systems \cite{tholence74,chen83}. The dashed line in Fig.~2b is the polynomial fit with $T_{0}=5$~K and $n=3.2$ and the dotted line represents a VF form, $\tau(T) = \exp{[E/k_{B}(T-T_{0})]}$ with $E=25$~K and $T_{0}=3.5$~K. For large relaxation times, both fits reproduce the data, yet the VF form falters when taken over the entire data range. The critical, power-law, fit describes the evolution of the characteristic charge dynamics over four decades in frequency.  At this stage, however, no functional form can be conclusively ruled out.
\begin{figure}[tbp]
\centering  \includegraphics[width=8.6cm,clip]{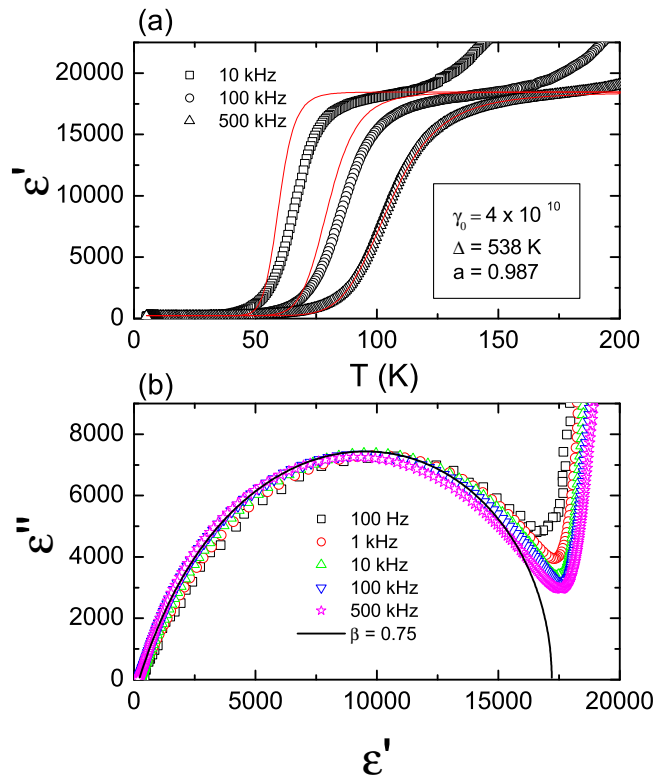}
\caption{(a) The real part $\varepsilon ^{\prime}$ of La$_{2-x}$Sr$_{x}$NiO$_{4}$ ($x=1/3$) is shown as a function of temperature for $E \parallel ab-$plane. Solid lines are best fits of the defect model. (b) The imaginary part $\varepsilon ^{\prime \prime}$ vs. the real part $\varepsilon ^{\prime}$. Solid line is the Cole-Cole arc with $\beta = 0.75$.}
\label{figure3}
\end{figure}

We note that the power law form $\tau (T) = a(T-T_{0})^{-n}$ is consistent with critical scaling wherein the exponent $n = \nu z$, with $\nu$ the standard spatial correlation length critical exponent $(\xi \sim (T-T_{0})^{-\nu})$ and a correlation time, $\tau \sim \xi^{z}$. If $T_{0}$ goes to zero as a function of parameters such as doping concentration or pressure, then the Davidson-Cole and similar expressions for glass formers would be naturally consistent with the very general quantum critical scaling form for the complex dielectric function $\varepsilon = \xi ^{-d_{M}} F(\omega \xi^{z}, \omega/T)$ in which $F$ has no dependence on its second argument and $d_{M}$ is a scaling exponent \cite{varma02}.

We have also studied Sr-doped lanthanum nickelate, La$_{2-x}$Sr$_{x}$NiO$_{4}$ (x=1/3). Fig.~3a shows the real part $\varepsilon ^{\prime}$ of the nickelate as a function of temperature at representative frequencies. The dielectric response is remarkably similar to that of La$_{2}$Cu$_{1-x}$Li$_{x}$O$_{4}$, i.e., large dielectric values at high temperatures and a sharp drop at low temperatures. The defect cell model \cite{ramirez02} accounts for these variations as in the cuprate. First, we fit Eq.~1 to the 500-kHz data and obtain a fit with $\gamma _{0}=4\times 10^{10}$, $a=0.987$, and $\Delta =538$~K. The relaxation rate $\gamma_{0}$ is one order of magnitude larger than that of the cuprate, indicating, within this framework, that the relevant defective cell in the nickelate has a smaller effective mass than the cuprate. The activation energy in the nickelate is also larger by a factor of 5, shifting $T_{f}$ to higher temperature relative to the cuprate. The solid lines in Fig.~3a are calculated with these same fitting parameters for each frequency. The overall fit is as good as that of the cuprate. The Cole-Cole arc for Sr-doped nickelate shows universal scaling (Fig.~3b) and is explained well by the Davidson-Cole expression (Eq.~2) with $\beta =0.75$.

Both the defect model \cite{ramirez02} and a charge-glass state account for the anomalous dielectric responses equally well, pointing toward underlying inhomogeneity. In the defect model where isolated defects are important, inhomogeneity depends on the particular dopant, its site, and concentration. We find very similar behaviors for Sr-substitutions on the La-site in La$_{2-x}$Sr$_{x}$NiO$_{4}$ and for Li-substitutions on the copper site in La$_{2}$Cu$_{1-x}$Li$_{x}$O$_{4}$. Further, the dopant concentration may be large enough, especially in La$_{2-x}$Sr$_{x}$Ni$_{2}$O$_{4}$, that a co-operative transition to a dipole glass state may occur and the dynamical enhancements to the dielectric constant at high temperature could disappear. Additionally, NMR studies of La$_{2}$Cu$_{1-x}$Li$_{x}$O$_{4}$ \cite{suh98} showed that the inhomogeneity is insensitive to the detailed nature of the individual dopant. Taken together, although the defect model captures the frequency and temperature dependence of the dielectric behavior, it is probably not applicable to understanding the fundamental origin of our observations.
 
Charge inhomogeneity also is established in Sr-doped La$_{2-x}$Sr$_{x}$NiO$_{4}$ where a charge-stripe glass was found for $x=0.20$ and $0.25$ \cite{hatton02}. In the cuprates, a similar charge-stripe glass was suggested as a precursor to cluster spin glass formation \cite{julien99,niedermayer98,suh98}. The correlation between spin and charge degrees of freedom also has been established  through the observation \cite{cao94} of an anomaly in the dielectric constant $\varepsilon ^{\prime}$ at the N$\acute{e}$el temperature $T_{N}$ of lightly hole-doped lanthanum cuprate. Fig.~4 shows the susceptibility $\chi$ of La$_{2}$Cu$_{1-x}$Li$_{x}$O$_{4}$ ($x=0.023$) (top panel) and La$_{1.66}$Sr$_{0.33}$NiO$_{4}$ (bottom panel) as a function of temperature. A deviation between zero-field cooled and field cooled data and a broad peak in ZFC $\chi$ are consistent with a spin glass state \cite{chou95} in both. The observed deviation temperatures are similar to the spin-glass temperatures determined by a scaling analysis \cite{chou95} and $\mu$SR measurement \cite{heffner02} and are comparable to the zero-frequency freezing temperatures obtained through the Davidson-Cole analysis, indicating a correlation between inhomogeneous magnetic and charge dynamics.
\begin{figure}[tbp]
\centering  \includegraphics[width=8.6cm,clip]{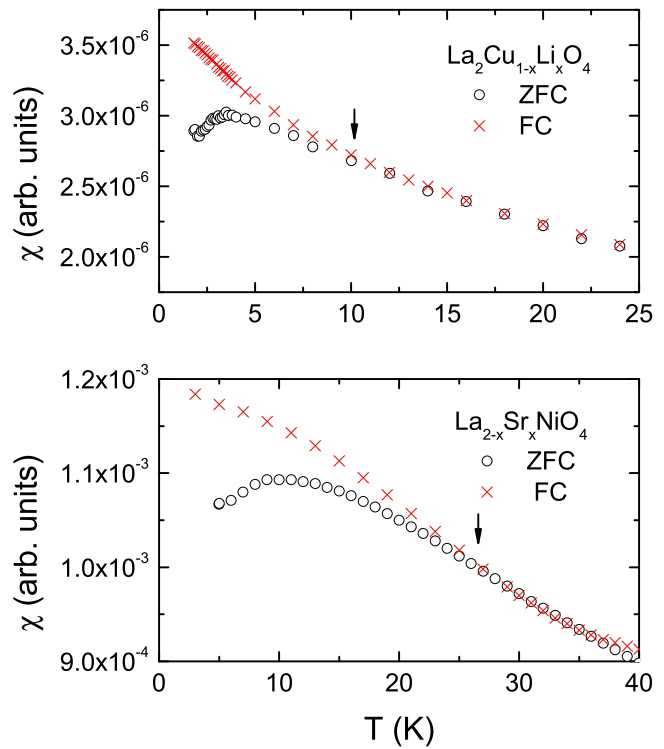}
\caption{Magnetic susceptibility $\chi$ vs. $T$ for La$_{2}$Cu$_{1-x}$Li$_{x}$O$_{4}$ ($x=0.023$) for $H \parallel c-$axis (top panel) and for La$_{2-x}$Sr$_{x}$NiO$_{4}$ ($x=1/3$) for $H \parallel ab-$plane (bottom panel). The FC data of the nickelate are lowered by a constant for comparison. The arrows indicate the temperature where a deviation occurs between ZFC (circles) and FC curves (crosses).}
\label{figure4}
\end{figure}

We emphasize that our observations are not specific to the existence of stripe glassiness. Stripes are not conclusively seen in Li-doped 214, albeit being omnipresent in the nickelates. Amongst others, it has been suggested that skyrmion-like patterns are triggered by the relatively immobile Li dopants, \textit{e.g.} \cite{haas96}. As the phenomena that we observe are common to both Li doped cuprates and to nickelates, stripes are not obviously the cause. Nevertheless, there is little doubt that both systems exhibit charge inhomogeneities (stripes in the nickelates and stripe/potential non-stripe patterns in the Li doped 214). These non-uniformities are common to both systems and drive the glassy dielectric behaviors we observe. The temperature scale on which charge glassiness develops correlates with the onset of glassy spin dynamics, suggesting their interrelationship. 

The work at Los Alamos was performed under the auspices of the U.S. Department of Energy. SWC was supported by the NSF-DMR-0103858. T. Park acknowledges benefits from discussion with J.-K. Lee and N. Curro.

$^{*}$ On leave from the Institute of High Pressure Physics, Russian Academy of Science, Troitsk.
\bibliography{glass}

\end{document}